\documentclass{kluwer}
\begin{document}
\begin{article}
\begin{opening}
\title{Identifying variable $\gamma$-ray sources through radio
observations 
}            

\author{J.M. \surname{Paredes}\email{jmparedes@ub.edu}}
\institute{Universitat de Barcelona}
\author{J. \surname{Mart\'{\i}}
} 
\institute{Universidad de Ja\'en}                               
\author{D.F. \surname{Torres}}
\institute{Lawrence Livermore National Laboratory}
\author{G.E. \surname{Romero}}
\institute{Instituto Argentino de Radioastronom\'{\i}a \\
FCAGLP, Universidad Nacional de la Plata}
\author{J.A. \surname{Combi}}
\institute{Universidad de Ja\'en} 
\author{V. \surname{Bosch-Ramon}}
\author{J. \surname{Garc\'{\i}a-S\'anchez}}
\institute{Universitat de Barcelona}




\runningtitle{Identifying variable $\gamma$-ray sources}
\runningauthor{Paredes et al.}

\begin{ao}
Josep M. Paredes\\
Departament d'Astronomia i Meteorologia\\
Facultat de F\'{\i}sica. Universitat de Barcelona \\
Av. Diagonal 647 \\
08028 Barcelona\\
Spain
\end{ao} 


\begin{abstract} 
We present preliminary results of a campaign undertaken with different
radio interferometers to observe
a sample of the most variable unidentified EGRET sources. 
We expect to detect which of the
possible counterparts of the $\gamma$-ray sources (any of the
radio emitters in the field) varies in time with similar timescales
as the $\gamma$-ray variation. If the $\gamma$-rays 
are produced in a jet-like source, as we have modelled theoretically,
synchrotron emission is also expected at radio wavelengths. Such radio emission
should appear variable in time and correlated with the $\gamma$-ray variability.

\end{abstract}




\end{opening}
\section{Introduction}
The Third EGRET Catalog (Hartman et al. 1999) lists 271 point sources. 
About two thirds of them have
no conclusive counterparts at lower frequencies. Even worse, 40 of
them do not show any positional coincidence (within the 95\% EGRET
contour) with possible $\gamma$-ray emitting objects known in our
galaxy (Romero, Benaglia, \& Torres 1999, Torres et al. 2001). In
order to understand the origin of all these unidentified
detections, their variability status is of fundamental importance.
Classic known models for $\gamma$-ray sources in our galaxy would
produce non-variable sources during the timescale of observations.
This is the case of pulsars (Thompson 2001), supernova remnants
in interaction with molecular clouds (Torres et al. 2003) and microquasar
jets in interaction with high density interstellar medium (Bosch-Ramon et
al. 2004a).
Alternatively, if some of the sources are produced by compact
objects, such as isolated magnetized black holes (Punsly et al.
2000) or microquasars (Paredes et al. 2000), high levels of flux 
variability can be expected.

\begin{figure} 
\vspace{8cm}
\includegraphics{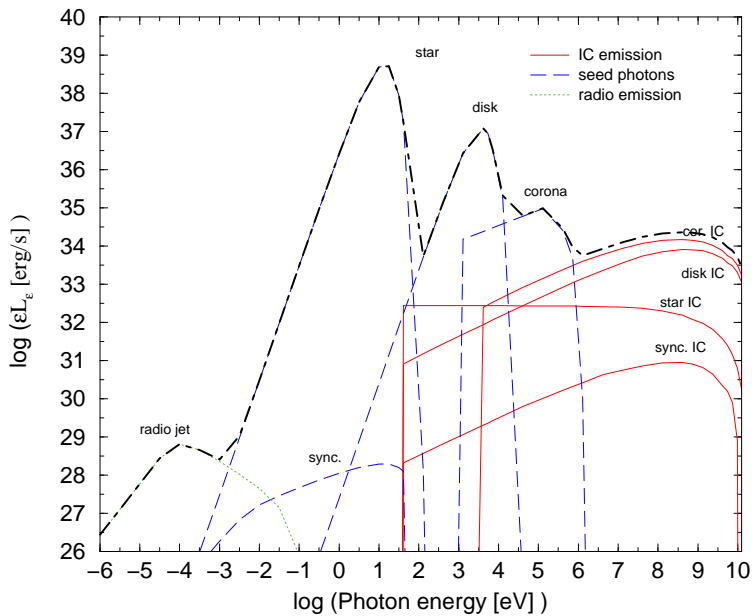}
\caption[]{Spectral energy distribution of a typical microquasar emitting up to EGRET energies (Bosch-Ramon et al. 2004b).}
\label{multi}
\end{figure}

According to recent theoretical models, the observed variable $\gamma$-ray 
emission could be associated with galactic compact objects such as 
X-ray binary  systems and isolated black holes. A model
considers the so-called microquasar and microblazar X-ray binaries,
in which the optical/UV photons from an optical star are inverse Compton 
up-scattered by relativistic electrons in the jets emanating from 
the accretion disc of a black hole/neutron star companion. Variability is
then naturally induced by the precesion of the jets, eccentricity of the orbit  and/or variable accretion rates. Moreover, it should be 
also detectable at radio wavelengths where synchrotron radio photons are
produced by the same relativistic electrons. A semi-analytical model
(see Fig.~\ref{multi}) of this and other processes, in a microquasar scenario,
is currently being developed by us with the purpose of
understanding the spectral energy distribution of EGRET sources across the whole
electromagnetic spectrum (Bosch-Ramon et al. 2004b). The non-thermal spectral
energy distribution has two peaks, one from the synchrotron jet
emission, at radio-IR energies, and other from Comptonization of
seed photons (stellar, coronal, disc and/or synchrotron photons), at MeV-GeV energies. The amplitude of the radio
variation would be somewhere between 10\%, as observed in the microquasar
LS5039/3EG~J1824$-$1514 (Rib\'o 2002, Rib\'o et al. 2004), and one order of magnitude, as
predicted in some microblazar models (Romero et al. 2002, Kaufman-Bernad\'o et al.
2002). The time scales could range between the orbital and the precession 
periods typical of X-ray binaries (from days to a few months).

Another possibility for a variable galactic $\gamma$-ray source is
that of an isolated black hole. It has been suggested that
such objects could radiate $\gamma$-rays due to spherical
accretion from the interstellar medium (Dermer 1997) or thanks to
a charged magnetosphere (Punsly 1998, Punsly et al. 2000). 
Finally, yet another possibility is the $\gamma$-ray emission
produced in the disc of a binary system formed by a pulsar and a
massive star. The pulsar accelerates charged particles in its
magnetosphere, which ultimately collide against the accretion disc
formed by the matter from the companion star. This collision yields hadronic (pp) interactions, and variability naturally results as a consequence
of the formation and loss of the accretion disc along the binary
orbit. A situation like this was explored in the case of A0535+26
by Romero et al. (2001), and was found able to explain the
observations for 3EG 0542+2610. In this case, however, the
$\gamma$-ray emission is mainly hadronic, and we do not expect
intense radio counterparts. 
All the previous models could produce variability indices in agreement with those determined by Torres et al. (2001). 

Identifying the most variable sources --and particularly those
which simultaneously show steep spectral indices at $\gamma$-rays -- is bound to
yield very interesting discoveries. In this work we report preliminary results of a search for variable radio counterparts of $\gamma$-ray variable EGRET sources using different radio interferometers.

\section{Selected sources to be studied}

From the original set of unidentified EGRET sources at
$|b|\leq 10^{\circ}$, a large fraction $\sim $ 50\% (40 sources) remains
without any known possible galactic counterpart (Romero, Benaglia,
\& Torres 1999). Since it is most unlikely that all these sources are extragalactic, they must 
encompass a population of variable galactic $\gamma$-ray sources yet to be 
discovered. 

To quantify the variability status of these sources we use the $I$
index, which establishes how variable a source is with respect to the
pulsar population. The reader is referred to Torres et al. (2001), and Torres, Pessah, \&
Romero (2001) for details and comparisons with other variability estimators
proposed in the literature. 
Sources with $I>2.5$ are already 3$\sigma$ away from the
statistical variability of pulsars, and we can be confident in
that its variability status might be indicative of its ultimate
nature. Selected EGRET sources that, being not in positional coincidence
with any known plausible galactic counterpart, are most likely
variable, are listed in Table~\ref{sources}. There, we give their 3EG
name, their galactic coordinates, the error in their position, in degrees
(assumed to be the 95\% confidence contours given by the 3EG
catalog), their gamma spectral index, 
$\Gamma$ (such that the photon distribution is given by $N(E)
\propto E^{-\Gamma}$), their EGRET flux and the index of variability $I$. 
In a first step, we included only those with smaller EGRET position error and/or without
hour angle limitation for the interferometer to be used.
Their range of variability index is 2.61 to 5.33 which can be considered as very significant. 
The typical size of the EGRET 95\% confidence contours is $\sim0.6^{\circ}$. 
It is also worth noticing
that the mean value of
the EGRET spectral indices is quite steep, steeper than
the steepest pulsar spectrum known.

\begin{table*}
\caption[]{Unidentified 3EG sources without known possible
galactic counterparts and visible from WSRT.}
\label{sources}
\begin{flushleft}
\begin{tabular}{l c c c c c c}
\hline \noalign{\smallskip} $\gamma$-Source &  $l$ & $b$ &
$\Delta\theta$ & $\Gamma$ & $\left< F_{\gamma}\right> \times
10^{-8}$ & 
$I $\\(3EG J)& (deg) & (deg) & & &(ph cm$^{-2}$ s$^{-1}$)  & \\
\noalign{\smallskip} \hline \noalign{\smallskip}



1735$-$1500  & 10.73 &    9.22 &   0.77 &$3.24 \pm 0.47$
&19.0 
&8.86 \\


1746$-$1001& 16.34 &    9.64 &    0.76 &$2.55 \pm 0.18$
&29.7 
&3.19 \\

1810$-$1032  & 18.81  &   4.23 &   0.39&$2.29 \pm 0.16$
 &31.5 
&2.61 \\

1812$-$1316 &16.70 &    2.39  &  0.39&$2.29 \pm 0.11$
 &43.0 
&2.60 \\

1828$+$0142 & 31.90 &    5.78 &   0.55&$2.76 \pm 0.39$
&30.8 
&5.33 \\

1834$-$2803&5.92 &   $-$8.97  &  0.52&$2.62 \pm 0.20$
&17.9 
&2.83 \\

1904$-$1124 & 24.22 &  $-$8.12  &  0.50&$2.60 \pm 0.21$
&22.5 
&2.91 \\

1928$+$1733 & 52.91  &   0.07  &  0.75&$2.23 \pm 0.32$
&38.6 
&3.99 \\

2035$+$4441 &83.17 &    2.50 &   0.54&$2.08 \pm 0.26$
&39.1 
&3.35 \\

\hline
\end{tabular}
\end{flushleft}
\end{table*}

The sample in Table~\ref{sources} 
was initially investigated for possible
radio counterparts (although no information on source variability
could be gathered at this stage). We constructed radio maps at the 20 cm wavelength using data from 
the NRAO VLA Sky Survey \footnote {The National Radio Astronomy Observatory
is a facility of the National Science Foundation operated under cooperative 
agreement by Associated Universities, Inc.}(NVSS, Condon et al. 1998) for each of 
these EGRET detections
and classified the radio sources within each EGRET confidence
contours. There are between 7 and 30 possible radio counterparts
(mostly uncatalogued) in each case that need to be monitored in
order to establish which of them --if any-- is varying in time.

The $\gamma$-ray variability of all our targets clearly
points out to a compact object as the origin of the high energy
emission. 
3EG~J1828+0142 has a previously
known possible extragalactic counterpart although outside the 95\%
EGRET confidence contour. It is an active
galactic nucleus referred to
as J1826+0149 by Sowards-Emmerd et al. (2003), who report it to be a flat spectrum radio
quasar. Halpern et al. (2003) also found this association plausible. 
However, an alternative galactic model also exists for this source
(Punsly et al. 2000, Butt et al. 2002, Combi et al. 2001, Bosch-Ramon et al. 2004c). In another case,
3EG~J1735$-$1500, a radiogalaxy within the location error box has been proposed as a counterpart 
although alternative candidates exist as well (Combi et al. 2003, Bosch-Ramon et al. 2004c). 

\section{Radio Observations}

In order to identify the origin of our selected $\gamma$-ray sources, 
we have undertaken a programme of radio observations using different interferometers. Most of our data has been obtained with the
Westerbork Synthesis Radio Telescope (WSRT) at the 21~cm wavelength.
In addition, the Very Large Array (VLA) of the NRAO was also used
for an initial test of our identification strategy. 
With this approach, we expect to detect which of the
possible counterparts of the $\gamma$-ray sources (any of the
radio emitters in the field) varies in time with similar timescales
to that of the $\gamma$-ray variation. If the $\gamma$-ray emission
is produced in a jet-like source, for instance, synchrotron
emission leading to radio detections, will also show a time
dependence. Flux densities of at least $\sim1$ mJy are expected by analogy with radio emitting
X-ray binaries.
The current status of our identification project is described in the following sections.

\subsection{WSRT multi-epoch observations}

The WSRT observations were conducted so far
for two of the targets in Table I, namely 3EG~J1928+1733 and 3EG~J2035+441.
We observed them on 2003, June 15 (WSRT1) and September 20 (WSRT2), 
thus with a time separation of three months.
In both cases we observed at the 21~cm wavelength, with 
a 20 MHz bandwitdth, a correlator setup of 64 channels, 4 polarizations and 8 bands.
The array of antennas was always in the same configuration in order to facilitate
variability studies, with 72~m of separation between the dishes of the two movable pairs of antennae.
Calibrator sources were observed one hour before and
after each observing run following the standard procedures at the WSRT. 

The first observing epoch was devoted to obtain a 21~cm mosaicked full track of 12~h duration
for each EGRET field with the idea of using it as a template for future variability studies. 
The different pointings, seven per EGRET field, were arranged
with a hexagonal packing covering a field of $\sim2^{\circ}$ squared.
In most cases, this should be
enough to fully cover the 95\% confidence contour of the
EGRET position.
The pointings were corrected
for the primary beam response before being combined into a mosaic. This process
increases significantly the noise in the map far from the central pointing positions. 
The sky positions 
covered by more than one pointing were weighted with the inverse square of the rms noise
before averaging and combining them. The resulting mosaic has a typical rms noise of
$\sim 0.2$ mJy beam$^{-1}$ over most of its solid angle. A circular restoring beam of
$45^{\prime\prime}$ was used in order to make easier the comparison with NVSS images.

During the second observing epoch, the two EGRET fields were mosaicked during a single 12~h track
with the same instrumental setup of the first epoch.
Sensitivity with the mosaicked snapshots
was, of course, lower but well enough
to control the variability
of already known NVSS sources as well as to look for new fainter radio variables in the field
(up to $\sim1$ mJy level).
Subtraction maps between the two epochs, in the image plane, were found to be a suitable
tool to look for possible variables in the field.

\subsection{VLA observations}

A Director Discretionary Time (DDT) observation was also conducted 
at the VLA as a test of our identification strategy
for this project. The target was 3EG~J1812$-$1316, whose field we mosaicked with four
pointings only during a single observing epoch (2003 February 10). The VLA data were taken
at the 20~cm wavelength in D configuration with the idea of comparing them
directly with the NVSS images. The corresponding results
are discussed below together with those from the WSRT.

\section{Results}

\begin{figure} 
\vspace{10cm}
\includegraphics{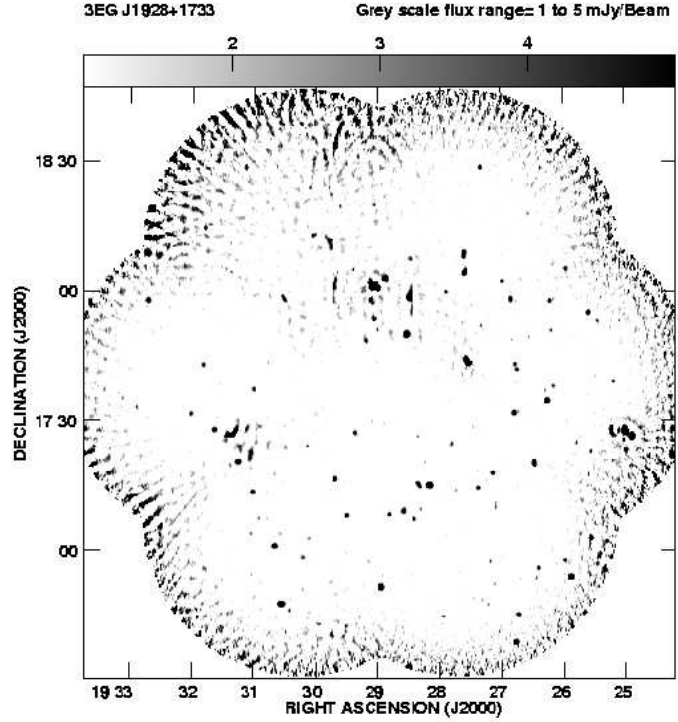}
\caption[]{WSRT mosaic of the 3EG~J1928$+$1733 field at 21 cm taken on 2003 June 15, the first of our two WSRT observing epochs.
A circular restoring beam of 45$^{"}$ has been used for easy comparison with the NVSS.}
\label{mos_1928}
\end{figure}

\begin{figure} 
\vspace{10cm}
\includegraphics{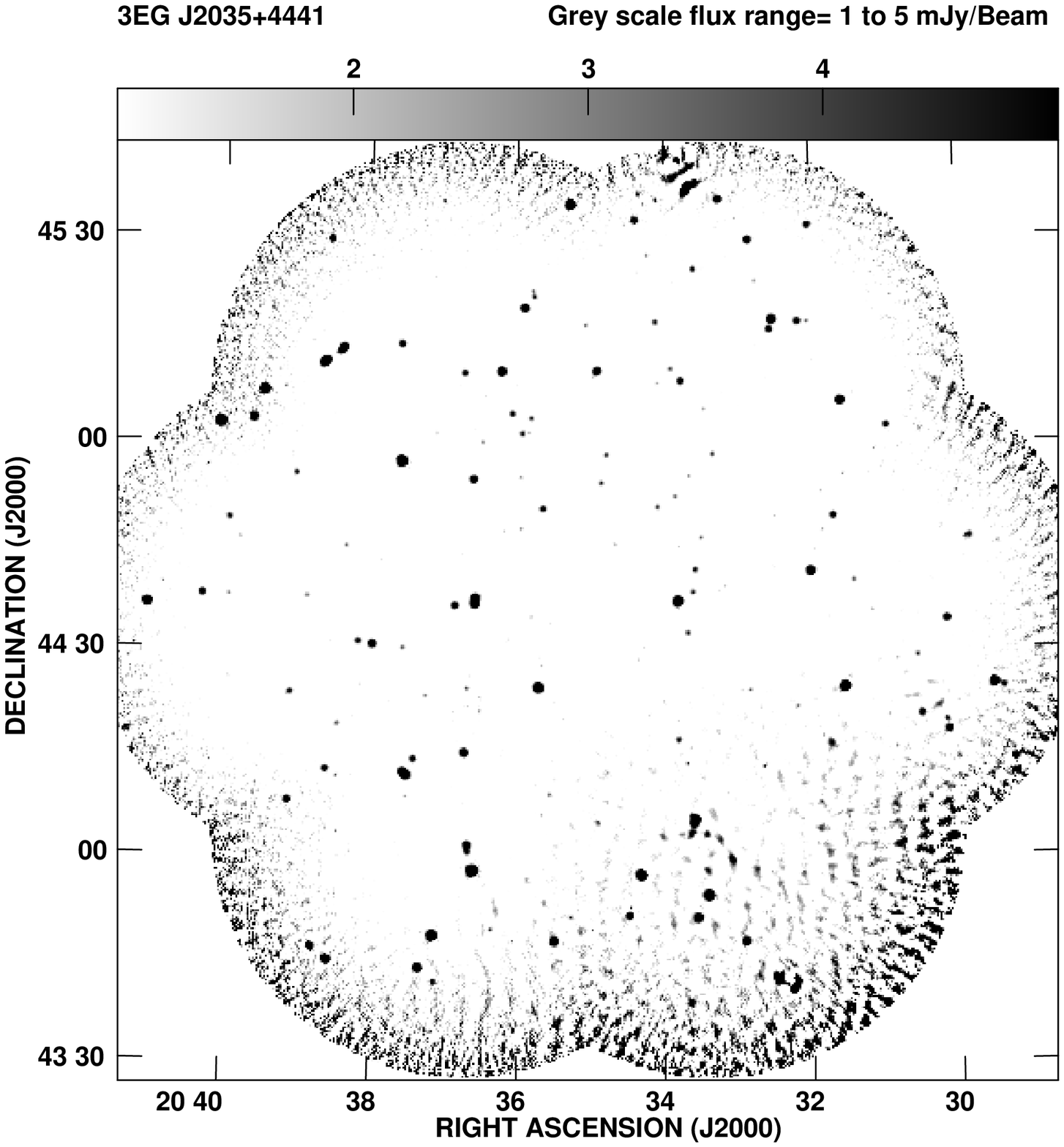}
\caption[]{WSRT mosaic of the 3EG~J2035$+$4441 field at 21 cm taken also on 2003 June 15. Nearly 100 radio sources are detected
in this image, but only  a few of them display possible evidence of variability. A 
circular restoring beam of 45$^{"}$ has been used as well.}
\label{mos_2035}
\end{figure}

Examples of our WSRT mosaics are presented in
Figs. \ref{mos_1928} and \ref{mos_2035}. Both correspond to
our first observing epoch and they have been computed using uniform weight.
Many radio sources are detected 
in each WSRT mosaic ($\sim100$) up to a brightness limit of 0.5 mJy. 

\begin{figure} 
\vspace{6cm}
\includegraphics{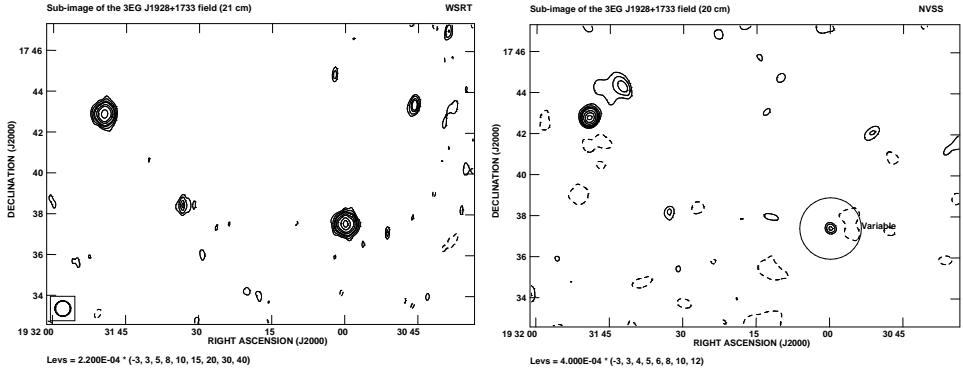}
\caption[]{Example of a clearly compact variable radio source discovered
when comparing our 3EG~J1928$+$1733 mosaic (left panel) with the NVSS (right panel). The proposed variable source, 
indicated with a circle, appeared significantly fainter a few years before our WSRT
observations.}
\label{zoom_1928}
\end{figure}

\begin{figure} 
\vspace{8cm}
\includegraphics{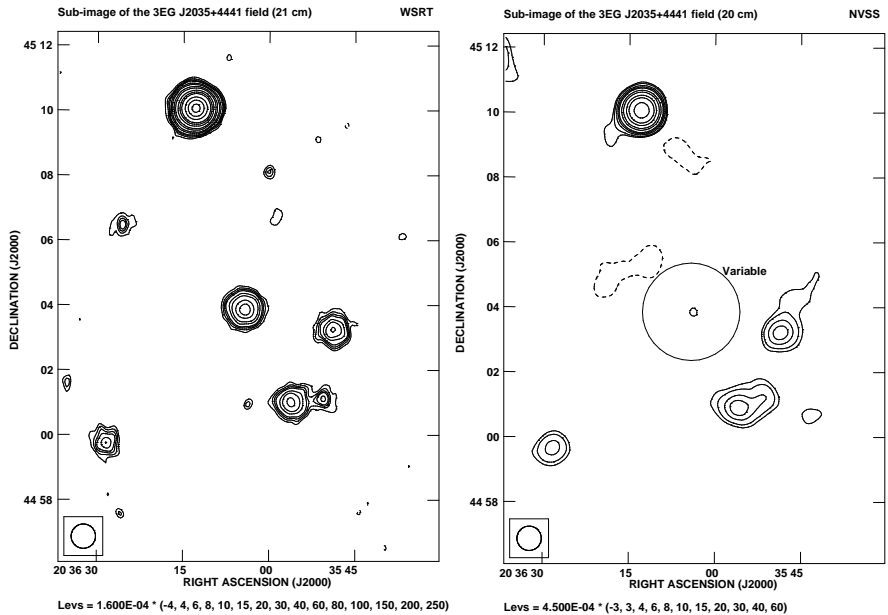}
\caption[]{Example of a clearly variable and compact radio source discovered
inside de 95\% confidence contour of the unidentified EGRET source 3EG~J2035$+$4441. 
The left panel is a zoomed region of a WSRT mosaic of the field as obtained by us
on 2003 June 15 at 21~cm. The right panel is the same region of the sky as it appeared in the
NVSS in 1995. The proposed variable source, indicated with a circle, was practically
undetectable some years ago.}
\label{zoom_2035}
\end{figure}

The comparison of our two WSRT epochs revealed initially no strongly flaring 
sources, at least in time scales of months. 
In Tables \ref{EG1928} and \ref{EG2035}, we list the WSRT positions and flux densities
of the possible radio variables detected (six in each field) 
as measured with the IMFIT task of the
AIPS package.  Some of the variable candidates 
reported here, with $\leq10$\% amplitude, are barely above the uncertainty in absolute flux calibration
and further observations are required to confirm them.
In spite of this, the comparison of our WSRT maps with 
the NVSS has actually revealed remarkable cases of apparent
radio transient, or flaring.
At least one radio transient source, in time scales of years, is present
among our variable candidates in both EGRET fields.
These interesting objects are marked
with an asterisk in Tables \ref{EG1928} and \ref{EG2035}, where the corresponding NVSS
flux density at the 20~cm wavelength is also included (upper limits are $3 \sigma$).
In addition, we illustrate them as well in Figs. \ref{zoom_1928} and \ref{zoom_2035}. 
The small wavelength difference, between the WSRT and the NVSS, is not likely to affect
the reality of the proposed radio transient sources for physically plausible spectral indices.
 
The comparison with the NVSS has been also very useful to identify
radio variable candidates in our single epoch VLA observations of 3EG~J1812$-$1316, again 
in time scales of years.  The results are presented in Table~\ref{EG1812} together
with the NVSS data. Two of the sources in the field, also marked with an asterisk, appear as reliable
variable radio sources. It is important to mention here that our VLA data was obtained in the
same D configuration of the array as the NVSS, hence comparison is straightforward.

Multiwavelength follow-up observations (X-ray, infrared,
optical, and eventually VLBI) of these so-detected confident 
candidates can ultimately yield to the discovery of new
microquasars or other $\gamma$-ray--emitting galactic compact
objects. We have initially inspected both the Digitized Sky Survey (DSS) and other databases
(e.g. 2MASS) looking for possible counterparts of our discovered radio variables.
The strong absorption in the Galactic plane at optical wavelengths makes the
search for an optical counterpart very difficult. In the infrared, the 2MASS
has evidence of a few sources having $K\sim$ 15 that could be consistent with some of
our $\pm$1$^{"}$ radio positions. 
 However, a confirmation with deep $K$-band images
is strongly necessary specially in cases where the 2MASS revealed no clear 
infrared source and a proposal for this goal has been recently submitted to the
Calar Alto 3.5~m telescope in Spain.
To determine the nature of the candidates, either galactic or extragalactic, infrared spectroscopy studies on the near infrared counterparts will be performed.
Similarly, the most interesting radio variable candidates have been 
proposed for Chandra observations in the X-ray band.

\begin{table}
\caption[]{Possible variable radio sources in the 3EG~J1928$+$1733 field. The most
reliable variables are marked with ($^{*}$).}
\label{EG1928}
\begin{flushleft}
\begin{tabular}{ccccc}
\hline \noalign{\smallskip} $RA (2000)$ &  $DEC (2000)$ & NVSS 
 & WSRT1  & WSRT2  \\
(h m s)& (~$^{\circ}$~~$^{'}$~~$^{"}$~)& (mJy) & (mJy) & (mJy)\\
\noalign{\smallskip} \hline \noalign{\smallskip}

19 27 35.56 $\pm$ 0.03 & $+$18 04 38.2 $\pm$ 0.6 & 38.1 $\pm$ 2.1 & 30.8 $\pm$
0.8 & 25.6 $\pm$ 1.1 \\
19 28 09.95 $\pm$ 0.01 & $+$17 15 17.4 $\pm$ 0.1 & 55.0 $\pm$ 1.7 & 62.8 $\pm$
0.6 & 61.0 $\pm$ 0.8 \\
19 30 31.40 $\pm$ 0.07$^a$ & $+$17 58 14.4 $\pm$ 0.7 & 5.2 $\pm$ 0.6 & 14.6 $\pm$ 0.6 &
22.8 $\pm$ 1.0 \\
19 31 00.91 $\pm$ 0.03$^{*}$ & $+$17 13 30.7 $\pm$ 0.5 & 2.5 $\pm$ 0.5 & 13.3 $\pm$
0.5 & 14.3 $\pm$ 0.8 \\
19 32 00.90 $\pm$ 0.05$^{*}$ & $+$17 31 44.4 $\pm$ 0.8 & $<$1.2  & 7.3 $\pm$
0.5 & 7.0 $\pm$ 0.8 \\
19 30 59.85 $\pm$ 0.04$^{*}$ & $+$17 37 32.7 $\pm$ 0.6 & $<$1.2  & 9.0 $\pm$
0.5 & 7.2 $\pm$ 0.7 \\
\hline
\end{tabular}
$^a$ {\small Extended object. Apparent variability possibly affected by strong primary beam correction.}\\ 
\end{flushleft}
\end{table}

\begin{table}
\caption[]{Possible variable radio sources in the 3EG~J2035$+$4441 field. The most
reliable variables are marked with ($^{*}$).}
\label{EG2035}
\begin{flushleft}
\begin{tabular}{ccccc}
\hline \noalign{\smallskip} $RA (2000)$ &  $DEC (2000)$ & NVSS 
 & WSRT1  & WSRT2  \\
(h m s)& (~$^{\circ}$~~$^{'}$~~$^{"}$~)& (mJy) & (mJy) & (mJy)\\
\noalign{\smallskip} \hline \noalign{\smallskip}

20 37 31.42 $\pm$ 0.01 & $+$44 11 35.8 $\pm$ 0.1 & 91.6 $\pm$ 3.3 & 92.5 $\pm$
0.4 & 87.5 $\pm$ 0.7 \\
20 36 34.34 $\pm$ 0.01$^a$ & $+$44 36 53.6 $\pm$ 0.2 & 80.6 $\pm$ 3.0 & 87.3 $\pm$
0.7 & 79.1 $\pm$ 0.7 \\
20 36 04.26 $\pm$ 0.05$^{*}$ & $+$45 03 52.8 $\pm$ 0.5 & $<$1.4 & 7.2 $\pm$ 0.3 &
7.9 $\pm$ 0.5 \\
20 39 55.62 $\pm$ 0.12 & $+$44 37 38.9 $\pm$ 1.2 & $<$1.4 & 2.3 $\pm$
0.3 & 2.0 $\pm$ 0.4 \\
20 30 57.85 $\pm$ 0.03 & $+$45 02 14.1 $\pm$ 0.3 & 7.1 $\pm$ 0.6 & 8.3 $\pm$
0.3 & 9.0 $\pm$ 0.5 \\
20 37 02.85 $\pm$ 0.25 & $+$44 57 29.3 $\pm$ 3.2 & $<$1.4 & $<$0.6 & 2.2 $\pm$
0.5 \\

\hline
\end{tabular}
$^a$ {\small Double source. Position of brighter component given.}
\end{flushleft}
\end{table}

\begin{table}
\caption[]{Possible variable radio sources in the 3EG~J1812$-$1316 field. The most
reliable variables are marked with ($^{*}$).}
\label{EG1812}
\begin{flushleft}
\begin{tabular}{cccc}
\hline \noalign{\smallskip} $RA (2000)$ &  $DEC (2000)$ & NVSS
 & VLA/DDT   \\
(h m s)& (~$^{\circ}$~~$^{'}$~~$^{"}$~)& (mJy) & (mJy)\\
\noalign{\smallskip} \hline \noalign{\smallskip}

18 13 52.65 $\pm$ 0.13$^{*}$ & $-$13 30 11.8 $\pm$ 3.5 & 6.4 $\pm$ 0.5 & 10.4 $\pm$
1.7 \\
18 14 32.83 $\pm$ 0.06 & $-$13 30 37.4 $\pm$ 1.8 & 13.7 $\pm$ 1.1 & 18.9 $\pm$
1.6 \\
18 14 26.88 $\pm$ 0.06$^{*}$ & $-$13 20 18.6 $\pm$ 1.8 & 12.0 $\pm$ 0.6 & 23.2 $\pm$
1.8 \\

\hline
\end{tabular}
\end{flushleft}
\end{table}

\acknowledgements
The Westerbork Synthesis Radio Telescope is operated by the ASTRON 
(Netherlands Foundation for Research in Astronomy) with support from the 
Netherlands Foundation for Scientific Research NWO.
J.M.P., V.B-R. and J.M. acknowledge partial support by DGI of the Ministerio de
Ciencia y Tecnolog\'{\i}a (Spain) under grant AYA-2001-3092, as well as
additional support from the European Regional Development Fund (ERDF/FEDER).
During this work, V.B-R. has been supported by the DGI of the Ministerio de
Ciencia y Tecnolog\'{\i}a (Spain) under the fellowship FP-2001-2699.
JM is also supported by the Plan Andaluz de Investigaci\'on (Spain) under project FQM322. The work of D.F.T. was performed under the
auspices of the US DOE (NNSA), by UC's LLNL under contract No.
W-7405-Eng-48. G.E.R. is supported by Fundaci\'on Antorchas and the 
Argentine Agencies CONICET and ANPCyT. This research has benefited from a cooperation agreement supported by Fundaci\'on Antorchas.

\end{article}
\end{document}